\documentclass[prb,twocolumn,showkeys,amsmath,amssymb]{revtex4-1}
\usepackage{graphicx}% Include figure files
\usepackage{dcolumn}% Align table columns on decimal point
\usepackage{bm}% bold math
\usepackage{tabularx}
\usepackage{color}
\usepackage{float}
\usepackage{physics}
\usepackage[mathcal]{euscript}
\begin{document}

\title{Realizing robust large-gap quantum spin Hall state in 2D HgTe monolayer on insulating substrate}

\author{Can Qi, Liying Ouyang, and Jun Hu}
\email[]{E-mail: jhu@suda.edu.cn}
\affiliation{School of Physical Science and Technology, Soochow University, Suzhou, Jiangsu 215006, China \\
Jiangsu Key Laboratory of Thin Films, Soochow University, Suzhou, Jiangsu 215006, China}

% Qi: adlison@foxmail.com; Ouyang: 825896778@qq.com
%\date{\today}% It is always \today, today,
             %  but any date may be explicitly specified

\begin{abstract}
Although many possible two-dimensional (2D) topological insulators (TIs) have been predicted in recent years, there is still lack of experimentally realizable 2D TI. Through first-principles and tight-binding simulations, we found an effective way to stabilize the robust quantum spin Hall state with a large nontrivial gap of 227 meV in 2D honeycomb HgTe monolayer by the Al$_2$O$_3$(0001) substrate. The band topology originates from the band inversion between the $s-$like and $p-$like orbitals that are contributed completely by the Hg and Te atoms, so the quantized edge states are restricted within the honeycomb HgTe monolayer. Meanwhile, the strong interaction between HgTe and Al$_2$O$_3$(0001) ensures high stability of the atomic structure. Therefore, the TI states may be realized in HgTe/Al$_2$O$_3$(0001) at high temperature.
\end{abstract}


%\pacs{66.30.J-, 71.55.Gs, 71.15.Nc}

\maketitle

\section{Introduction}

Great interest in topological insulators (TIs) has been inspired in recent years, because they exhibit intriguing quantum spin Hall (QSH) states. \cite{KaneMele1, ZhangBernevig1, HgTeTheory,  Nagaosa, HgTeExp, KaneReview, QiReview} A QSH state is characterized by the combination of insulating bulk state and quantized helical conducting edge or surface state. Here the helical conducting state provides intrinsic spin lock and is robust against elastic backscattering and localization, so that the ITs are ideal for various applications that require dissipationless spin transport. \cite{KaneReview, QiReview} Although QSH state was firstly predicted in graphene which is an ideal two-dimensional (2D) material, it is difficult to observe it in graphene due to the weak intrinsic spin-orbit coupling (SOC). \cite{Hu-1, Hu-2, Hu-3} On the contrary, significant progress in experiment on the investigation of TIs has been made in three-dimensional (3D) TIs \cite{BiSb-theory, BiSb-exp, Bi2Se3-theory, Bi2Se3-exp1, Bi2Se3-exp2} and quantum wells \cite{HgTeTheory, HgTeExp, QW-1, QW-2, QW-3}. Nevertheless, 2D TIs are more suited for applications, because bulk carriers that often plague the 3D TIs can be vacated simply by gating. So far, many possible 2D TIs have been predicted, \cite{Ju-Li, WengHM1, Bansil-1, HeineT, ZhaoMW} but very few experimental achievements of 2D TIs have been reported, because most predicted 2D TIs need to be decorated by anion atoms (such as H and halogen elements) and keep freestanding. However, it is difficult to grow freestanding 2D TIs in experiment. For example, silicene was predicted to be TI with nontrivial gap of 1.55 meV, but the honeycomb silicene has been fabricated only on surfaces of transition metals such as Ag(111) and Ir(111). \cite{Silicene-1, Silicene-2, Silicene-3} Unfortunately, there is no evidence yet that silicene on transition metal surface retains the QSH state. Therefore, exploring 2D TIs is still a big challenge in both theoretical and experimental investigations.

To obtain experimentally realizable 2D TIs, two factors should be considered. Firstly, a 2D TI is usually grown on a substrate and contact with substrate as well in applications. In general, the 2D TIs and the substrate will be bound by chemical bonds between them, so that the structures are stable. Secondly the QSH state must be preserved on the substrate. Furthermore, to avoid electronic contamination of trivial carriers from substrate to the QSH state, insulators with large band gap are preferred to be the substrate. Recently, it was predicted that stanene grown on some substates shows large nontrivial band gap, but the top Sn atoms need to be iodinated. \cite{Sn-on-sub-1, WangHui} Zhou $et~al$ predicted that Bi or W monolayer with artificial honeycomb lattice on halogen decorated Si(111) surface also exhibits large nontrivial band gap. \cite{sd2-1, sd2-2} However, these systems may be difficult to be produced in experiment, because of the delicate atomic structures.

For a chemically bound system with a 2D TI and an insulating substrate, the strong hybridization mainly occurs between the vertically aligned orbitals, such as the $s$ and $p_z$ orbitals. Therefore, the QSH state will be destroyed if it originates from the $p_z$ orbital such as in graphene and silicene. In contrast, the in-plane $p_x$ and $p_y$ orbitals do not interact strongly with the substrate. Accordingly, the QSH state may be preserved if it derives from the $p_x$ and $p_y$ orbitals. Interestingly, the honeycomb HgTe and HgSe monolayer (ML) are possible candidates. Recently, it was predicted that the freestanding honeycomb HgTe and HgSe ML exhibit large TI gap at $\bf\Gamma$ point. \cite{HgTe-ML-1, HgTe-ML-2} However, it is difficult to produce freestanding honeycomb HgTe and HgSe ML, similar to the situation of silicene. It is known that the bulk HgTe and HgSe crystallize zinc blende structure, so the HgTe(111) surface with two atomic layers forms buckled honeycomb lattice. Clearly, if there is proper substrate, it is possible to grow ultrathin HgTe(111) film with only two atomic layers --- the honeycomb HgTe and HgSe ML. In fact, honeycomb ZnO and GaN ML have been fabricated in experiment recently, \cite{ZnO, GaN} which ensures the possibility to achieve honeycomb HgTe and HgSe ML. The $\alpha-$Al$_2$O$_3$(0001) surface has commensurate lattice with the honeycomb HgTe and HgSe ML and it is an insulator with large band gap, so it may be used as the substrate to grow the honeycomb HgTe and HgSe ML [notated as HgTe/Al$_2$O$_3$(0001) and HgSe/Al$_2$O$_3$(0001)]. Therefore, it is instructive to reveal whether HgTe/Al$_2$O$_3$(0001) and HgSe/Al$_2$O$_3$(0001) possess QSH states.

In this paper, we studied the structural and electronic properties of HgTe/Al$_2$O$_3$(0001) and HgSe/Al$_2$O$_3$(0001) by using first-principles calculations and tight-binding simulations. We found that the combination of  honeycomb HgTe ML and Al$_2$O$_3$(0001) can not only stabilize the atomic structure but also maintain the QSH state with a large TI gap of 227 meV, while HgSe/Al$_2$O$_3$(0001) becomes a trivial insulator with a gap. The band topology originates completely from the honeycomb HgTe ML, hence the quantized helical states are confined at the edges of the honeycomb HgTe ML. Together with the highly stable atomic structure, the QSH effect is expected to be realized in HgTe/Al$_2$O$_3$(0001) at high temperature.

\section{Methods}

To model the Al$_2$O$_3$(0001) substrate, we constructed a slab of 18 atomic layers with thickness of $\sim$11 {\AA}, and inserted a vacuum of 15 {\AA} between adjacent slabs. The structural and electronic properties were calculated with density functional theory (DFT) as implemented in the Vienna {\it ab-initio} simulation package. \cite{VASP1, VASP2} The interaction between valence electrons and ionic cores was described within the framework of the projector augmented wave (PAW) method. \cite{PAW1,PAW2} The fully occupied semicore $5d$ orbital of Hg is treated as valence orbital. The spin-polarized local density approximation (LDA) was used for the exchange-correlation potentials and the effect of SOC was invoked self-consistently. \cite{LDA} The energy cutoff for the plane wave basis expansion was set to 500 eV. The 2D Brillouin zone was sampled by a 36$\times$36 k-grid mesh. The atomic positions were fully relaxed with a criterion that requires the force on each atom smaller than 0.01 eV/{\AA}.

The band topology of a system is characterized by the topological invariant $\mathbb{Z}_2$, with $\mathbb{Z}_2=1$ for TIs and $\mathbb{Z}_2=0$ for ordinary insulators. \cite{Z2} For the systems without inversion symmetry, $\mathbb{Z}_2$ can be expressed with the Berry connection and the Berry curvature as
\begin{equation}
\mathbb{Z}_2 = \frac{1}{2\pi} \qty[\oint_{\partial \mathcal{B}^-} d\textbf{k}\cdot\textbf{A}(\textbf{k}) - \int_{\mathcal{B}^-}d^2k\mathcal{F}(\textbf{k})]~\mathrm{mod}~2,
\end{equation}
where $\textbf{A}(\textbf{k})=i\sum_n \braket{u_n(\textbf{k})}{\grad_k u_n(\textbf{k})}$ is the Berry connection and $\mathcal{F}(\textbf{k})=\grad_k \cross \textbf{A}(\textbf{k})|_z$ is the Berry curvature; $u_n(\textbf{k})$ is the the periodic part of Bloch function and the sum is over occupied bands. $\mathcal{B}^-$ and $\partial \mathcal{B}^-$ indicate half of 2D Brillouin zone and its boundary, respectively. The nonzero $\mathbb{Z}_2$ invariant is an obstruction to smoothly defining the Bloch functions in $\mathcal{B}^-$ under the time-reversal constraint. In DFT calculations, it is convenient to adopt the so-called $n-$field scheme to calculate the integration in Eq. (1). \cite{n-field-0, n-field-1, n-field-2}. For an $N \cross M$ $k-$grid mesh in the 2D Brillouin zone, the intervals along the two reciprocal vectors are $\bm{\mu}$ and $\bm{\nu}$, and $\bm{k}_j$ is the vector of a grid point. Then the integer field $n(\textbf{k}_j)$ can be defined as 
\begin{equation}
n(\textbf{k}_j) = \frac{1}{2\pi} \qty{[\Delta_{\bm{\nu}} \textbf{A}_{\bm{\mu}}(\textbf{k}_j) - \Delta_{\bm{\mu}} \textbf{A}_{\bm{\nu}}(\textbf{k}_j)] - \mathcal{F}(\textbf{k}_j)},
\end{equation}
where $\Delta_{\bm{\nu}}$ is the forward difference operator. The $\mathbb{Z}_2$ invariant is given by the summation of the $n$ field in $half$ of the Brillouin zone $\mathcal{B}^-$, i.e., $\mathbb{Z}_2 = \sum_{\textbf{k}_j \in \mathcal{B}^-}n(\textbf{k}_j)$ mod 2.

For a 2D TI, the band structure of its one-dimensional (1D) nanoribbon usually exhibits edge bands within the gap of the 2D TI. We used the tight-binding model as implemented in PythTB code \cite{PythTB}. In general, the Hamiltonian in spin space can be written as
\begin{equation}
H=\mqty(H_{\uparrow\uparrow} & H_{\uparrow\downarrow} \\ H_{\downarrow\uparrow} & H_{\downarrow\downarrow}),
\end{equation}
where $\uparrow$ and $\downarrow$ stand for the up and down spins, respectively. If the SOC effect is not included, $H_{\uparrow\downarrow} = H_{\downarrow\uparrow} = 0$. For a given basis set $\qty{C_{i\alpha}}$ with $i$ and $\alpha$ respectively indicating the orbital and spin, the matrix elements in Eq. (3) can be written as
\begin{equation}
H_{\alpha\beta} = -\sum_{\langle i,j \rangle} ( t_{i\alpha,j\beta} C_{i\alpha}^{\dagger} C_{j\beta} + H.C.),
\end{equation}
where $t_{i\alpha,j\beta}$ is the hopping parameter. When $\alpha \ne \beta$, $t_{i\alpha,j\beta}$ represents the spin-flip hopping induced by the SOC effect. By using the Wannier90 code, \cite{Wannier-3} the maximally localized Wannier functions (MLWFs) \cite{Wannier-1, Wannier-2} can be obtained based on the DFT calculations. Then, the hopping parameter between any two orbitals can be calculated with the MLWFs.

\section{Results and discussions}

\begin{table}[b]
 \centering
 \caption{The structural properties of freestanding (FS) and Al$_2$O$_3$(0001) supported (type-I and type-II) honeycomb HgTe ML. Different bond lengths (in \AA), the buckling distance (in \AA) and the inner angles (in degree) are indicated in Fig. 1 and 3. The binding energy is calculated as $E_b=E(HgTe) + E[Al_2O_3(0001)] - E[HgTe/Al_2O_3(0001)]$.
}
 \tabcolsep0.11in             % column seperation
 \begin{tabular}{cccccccccc}
   \hline
   \hline
  % after \\: \hline or \cline{col1-col2} \cline{col3-col4} ...
    & $d_1$ & $d_2$ & $d_3$ & $\Delta h$ & $\theta$ & $E_b$ \\
   \hline
   FS & 2.80 & & & 0.48 & 116.9 & \\
   Type-I & 2.87 & 2.74 & 2.70 & 0.84 & 111.9 & 0.99 \\
   Type-II & 2.78 & 3.54 & 3.46 & 0.42 & 117.8 & 0.04 \\
   \hline
   \hline
 \end{tabular}
\end{table}

\begin{figure}
\centering
\includegraphics[width=8.5cm]{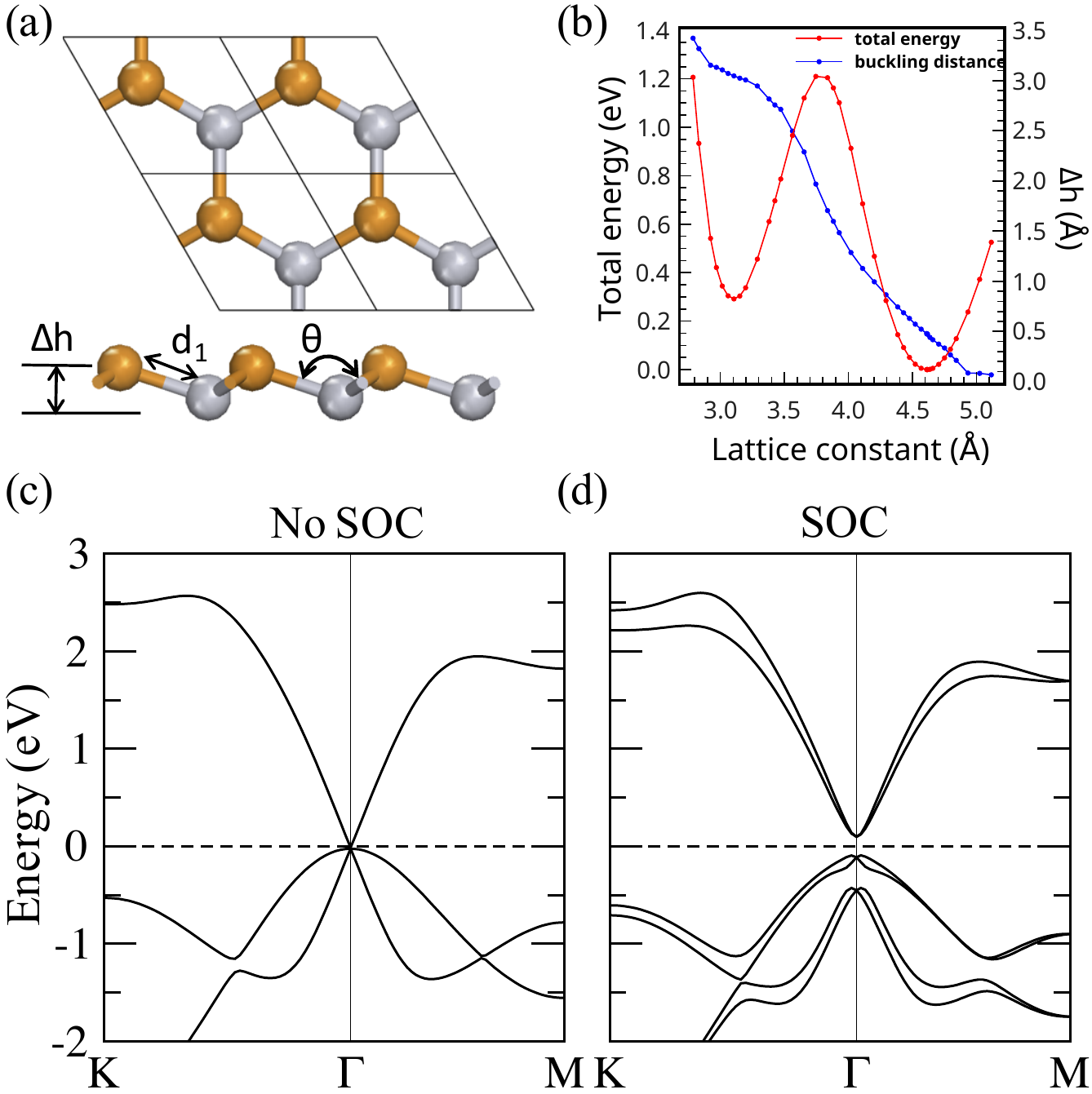}
\caption{(Color online) Properties of 2D honeycomb HgTe ML. (a) Top and side views of the atomic structure. The gray and orange spheres represent Hg and Te atoms, respectively. (b) Total energy and buckling distance ($\Delta$h) as a function of lattice constant. (c, d) Band structures without and with SOC, respectively. The horizontal dashed lines represent the Fermi level.
}\label{ML-2D}
\end{figure}

We first investigated the structural and electronic properties of the freestanding honeycomb HgTe ML which can be regarded as the HgTe(111) surface with only two atomic layers as shown in Fig. \ref{ML-2D}(a). To obtain the optimal lattice constant, we calculated the total energy as a function of lattice constant as plotted in Fig. \ref{ML-2D}(b). It can be seen that the ground state locates at $a=4.61$ {\AA} which is slightly larger than the lattice constant of HgTe(111) ($a=4.57$ \AA). The honeycomb HgTe ML keeps buckled with buckling distance of 0.48 \AA and Hg-Te bond length of 2.8 \AA as listed in Table I. This is different from other zinc blend semiconductors whose ultra-thin (111) surface prefers planar honeycomb structure. \cite{zincblend-111} Then we calculated the band structures without and with considering the SOC effect, respectively. As displayed in Fig. \ref{ML-2D}(c) and \ref{ML-2D}(d), without considering the SOC effect, the honeycomb HgTe ML is a semimetal with a Dirac cone at the $\vb\Gamma$ point. When the SOC is involved, a large band gap of 209 meV is induced at the $\vb\Gamma$ point.

\begin{figure}
\centering
\includegraphics[width=8.5cm]{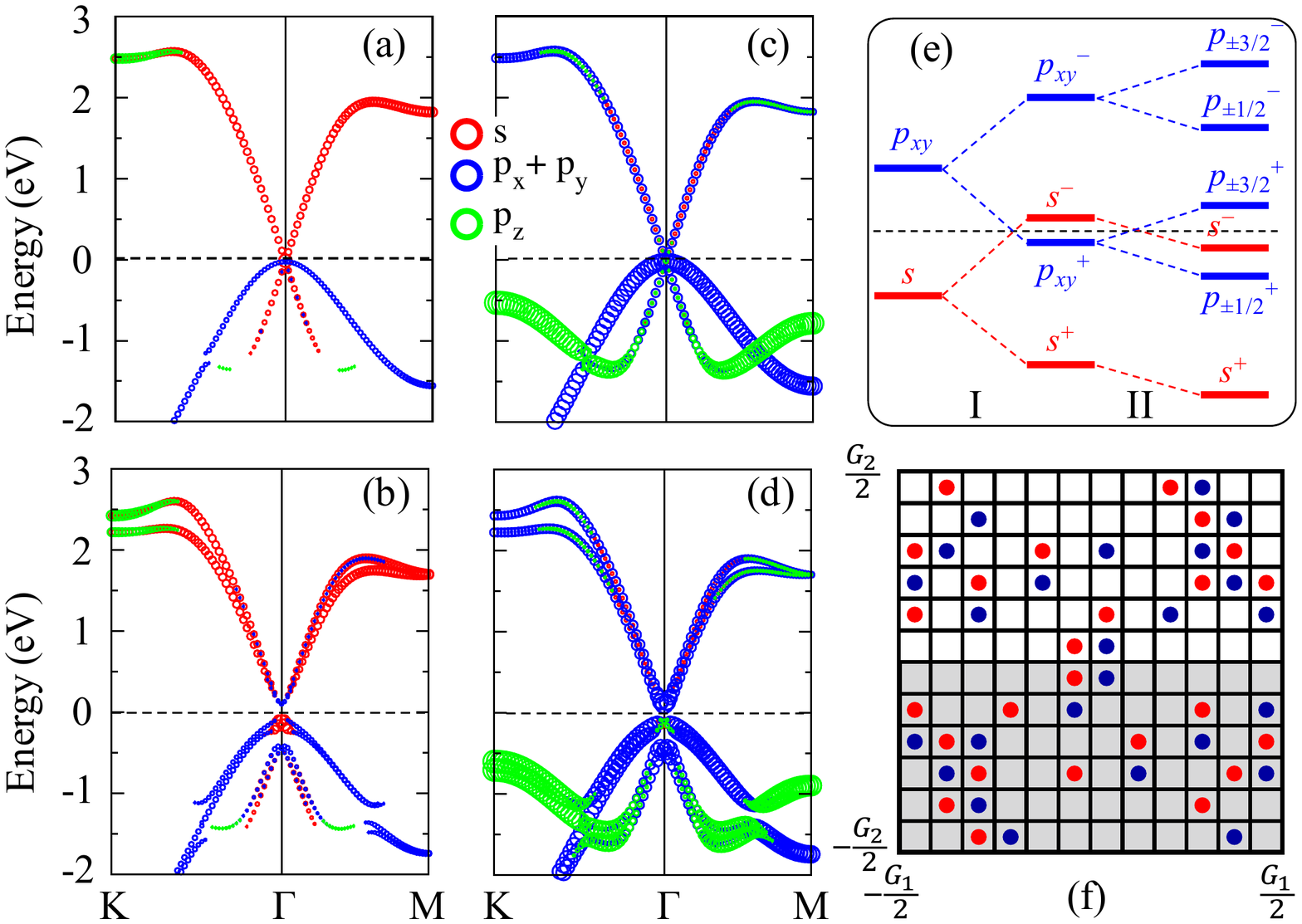}
\caption{(Color online) (a$-$d) Atomic orbital resolved band structures of the freestanding honeycomb HgTe ML without (top panels) and with (bottom panels) the SOC effect, projected on Hg  (a, b) and Te (c, d), respectively. The sizes of the circles indicate the weights of the corresponding atomic orbitals. The horizontal dashed lines represent the Fermi level. (e) Schematic diagram of the evolution of the $s$ and $p_{x,y}$ orbitals at $\vb\Gamma$ point. `I' and `II' denote the chemical bonding and SOC effect, respectively. The `$+$' and `$-$' signs indicate the bonding and antibonding orbitals. (f) The $n-$field configuration in the Brillouin zone spanned by G$_1$ and G$_2$. The nonzero points are denoted by red ($n=1$) and blue ($n=-1$) dots, respectively.
}\label{ML-level}
\end{figure}
 
To reveal the orbital characters of the states near the Fermi level ($E_F$) at the $\vb\Gamma$ point, we projected the band structures onto the atomic orbitals of the Hg and Te atoms. Note that the $p_x$ and $p_y$ orbitals are degenerate (notated as $p_{xy}$) due to the local $C_{3v}$ symmetry. As shown in Fig. \ref{ML-level}(a$-$d), the $p_z$ orbital of both Hg and Te atoms does not contribute to states near $E_F$ at the $\vb\Gamma$ point. In fact, the Dirac-like bands around the $\vb\Gamma$ point are dominated by the Hg$-s$ and Te$-p_{x,y}$ orbitals, while the parabolically dispersive band is mainly from the Te$-p_{x,y}$ orbital. To elucidate the evolution of the atomic orbitals, we extracted the energy levels at the $\vb\Gamma$ point and plotted in Fig. \ref{ML-level}(e). The chemical bonding between the Hg and Te atoms turns $s$ and $p_{x,y}$ into the bonding orbitals ($\ket{s^-}$, $\ket{p_{xy}^-}$) and antibonding orbitals ($\ket{s^+}$, $\ket{p_{xy}^+}$). It is clear that $\ket{s^-}$ and $\ket{p_{xy}^+}$ contribute to the bands near the Fermi level ($E_F$). In the absence of SOC, the energy level of $\ket{s^-}$ locates above that of $\ket{p_{xy}^+}$, resulting in a normal band order. After the SOC effect is taken into account, $\ket{p_{xy}^{\pm}}$ further splits into $\ket{{p_{\pm3/2}}^{\pm}}$ and $\ket{{p_{\pm1/2}}^{\pm}}$, with the former shifting upwards and the later downwards. Moreover, $\ket{s^-}$ and $\ket{p_{xy}^+}$ couple with each other through the SOC Hamiltonian, which reduces the energy level of $\ket{s^-}$. As a consequence, the energy levels of $\ket{s^-}$ and $\ket{p_{xy}^+}$ invert at $\vb\Gamma$ point, which implies nontrivial band topology. Accordingly, we calculated the $\mathbb{Z}_2$ invariant with the $n-$field method. By summing the positive and negative $n$ field numbers in half of the Brillouin zone [shadow area in the inset in Fig. \ref{ML-level}(f)], we obtained $\mathbb{Z}_2=1$, clearly demonstrating the nontrivial band topology of the honeycomb HgTe.

\begin{figure}
\centering
\includegraphics[width=8.5cm]{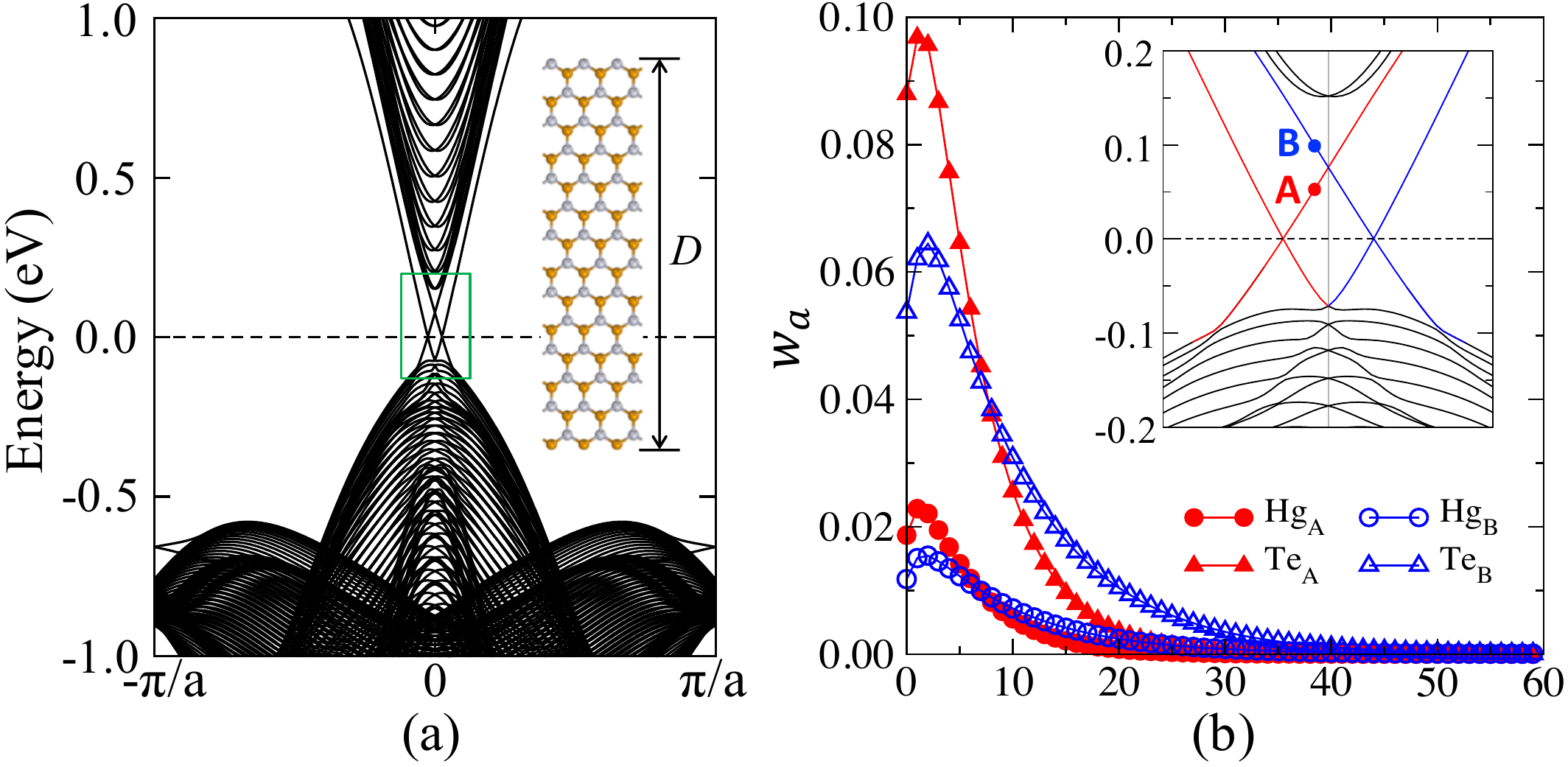}
\caption{(Color online) (a) Band structure of 1D HgTe nanoribbon with zigzag edges. The inset shows the structure of nanoribbon with width of 14 unit cells ($D=14~u.c.$). One edge is composed of Hg atoms, while the other edge of Te atoms. (b) Spatial distributions of the wavefunctions at A and B. Horizontal axis indicates the indexes of unit cells. $w_a$ is the weight of each atom. The inset is the band structure in the green rectangle in (a). The red and blue bands stand for spin-up and spin-down states, respectively.
}\label{ML-edge}
\end{figure}

It is known that the TIs are manifested by quantized edge states which can be seen in the band structure of the corresponding 1D nanoribbons. Therefore, we applied tight-binding model to calculate the band structure of a 1D HgTe nanoribbon. We should point out that if only the low-energy bands near $E_F$ are considered, a four-band model with only the first nearest neighboring hopping between the $s$ and $p_{xy}$ orbitals of Hg and Te is a good approximation. In our calculation, however, we considered all the hopping with $\abs{t_{i\alpha,j\beta}}>0.001$ eV, which includes interactions up to the sixth nearest neighbors. The hopping parameters have been tested rigorously, and the tight-binding model precisely reproduces the DFT band structures (see Fig. S1). A 1D HgTe nanoribbon with zigzag edges and width of 60 unit cells ($D=60~u.c.$) was chosen, and its band structure is plotted in Fig. \ref{ML-edge}(a). Apparently, linearly dispersive bands appear in the gap of the 2D honeycomb HgTe ML. Since a wavefunction is the linear combination of all the atomic orbitals with coefficient $C_{i\alpha}$, $\sum_{i\alpha} \abs{C_{i\alpha}}^2$ over the orbitals of a single atom gives the atomic weight $w_a$ from which we can obtain the spatial distributions of the wavefunctions. As shown in Fig. \ref{ML-edge}(b) for the `A' and `B' points, the wavefunctions of both bands are localized in the edge region on the same side, and the Te$-p_{xy}$ orbitals dominate the wavefunctions. From $C_{i\alpha}$, we found that the bands crossing `A' and `B' carry the spin-up and spin-down electrons, respectively. In addition, these bands have opposite slopes, which indicates that the spin-up and spin-down electrons propagate along opposite directions on the edge. Similarly, the other two bands (red and blue) in the inset in Fig. \ref{ML-edge}(b) carry the spin-up and spin-down electrons on the other side of the HgTe nanoribbon. Therefore, these edge bands provide the helical edge states for the QSH effect.

\begin{figure}
\centering
\includegraphics[width=8.5 cm]{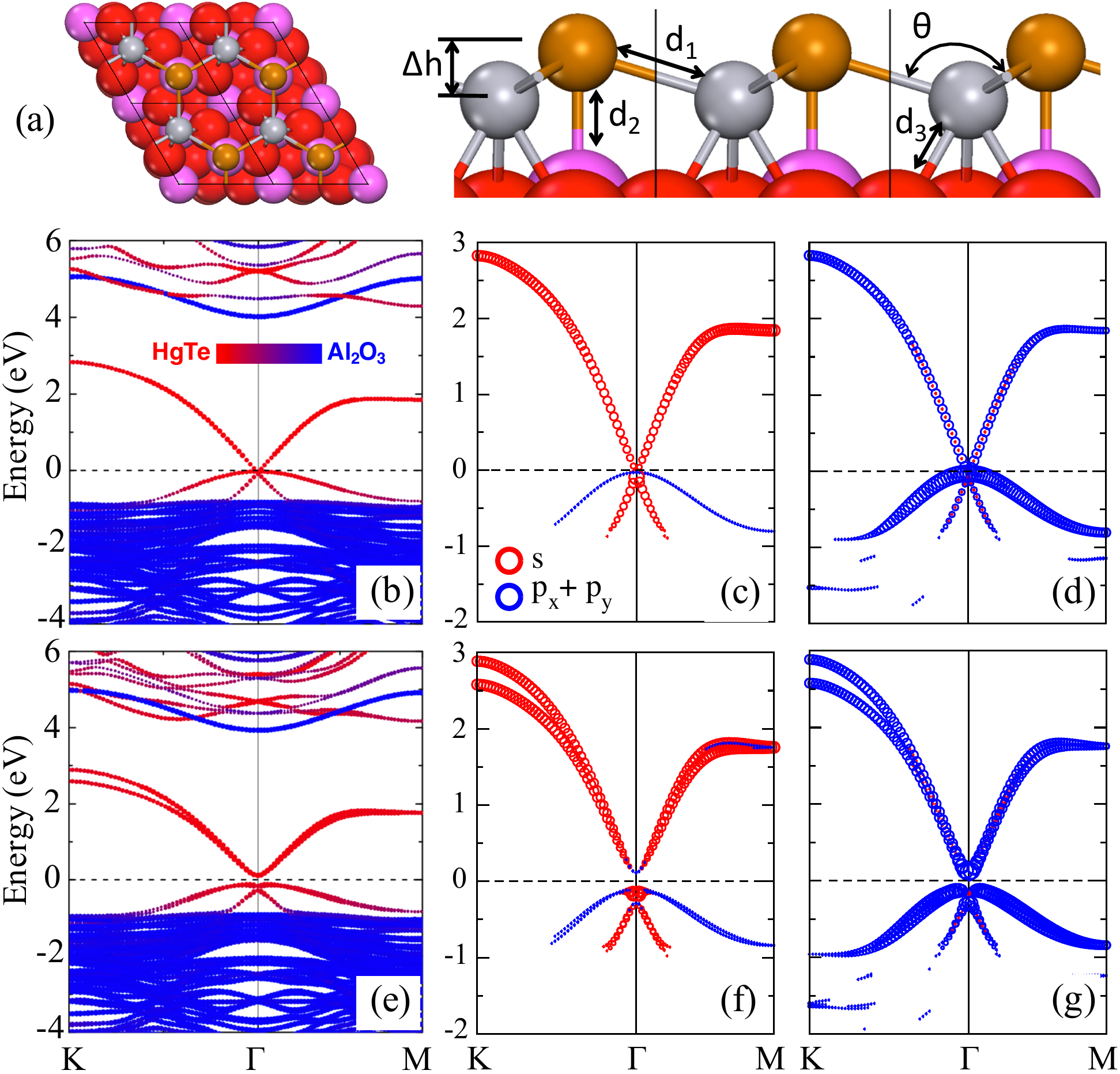}
\caption{(Color online) Atomic and electronic band structure of the type-I HgTe/Al$_2$O$_3$(0001). (a) The top and side views of the atomic structure. The gray, orange, purple and red spheres represent Hg, Te, Al and O atoms, respectively. (b-f) Band structures without (middle panels) and with (bottom panels) SOC. (b, e) indicate the contributions from the HgTe ML and Al$_2$O$_3$(0001) substrate represented by the color bar. (c, f) and (d, g) show the contributions of the atomic orbitals of the Hg and Te atoms, respectively. The size of the circles indicates the weights of the corresponding atomic orbitals to the bands.%The points with weight smaller than 8\% are ignored.
}\label{sub-band}
\end{figure}

Then we put the honeycomb HgTe ML on Al$_2$O$_3$(0001) ($a=4.76$ \AA) which imposes extensile strain of 3.2\% on the honeycomb HgTe ML. There are two adsorption configurations for the honeycomb HgTe ML. Fig. \ref{sub-band} shows the type-I configuration: the Hg atom locates over the center of three surface O atoms and the Te atom is above the surface Al atom. Switching the positions of the Hg and Te atoms yields the type-II configuration. It can be seen from Table I that the Hg-Te bond lengths in both configurations do not change much from those of the freestanding honeycomb HgTe ML. The binding between the HgTe and Al$_2$O$_3$(0001) is strong in the type-I configuration, resulting in short O-Hg and Al-Te bond lengths, higher buckling distance (0.84 \AA), and large adsorption energy (0.99 eV), as listed in Table I. The honeycomb HgTe ML is physisorbed on Al$_2$O$_3$(0001) in the type-II configuration, so that the O-Te and Al-Hg bond lengths are large and the adsorption energy is small (0.04 eV). Therefore, it is most probable to obtain the type-I configuration when growing honeycomb HgTe ML on Al$_2$O$_3$(0001).

It is interesting to see how the electronic structure of the honeycomb HgTe ML changes with the interaction from Al$_2$O$_3$(0001) substrate. Hence we calculated the band structure of the type-I HgTe/Al$_2$O$_3$(0001). It can be seen from Fig. \ref{sub-band}(b) and \ref{sub-band}(e) that the bands locating in the gap of Al$_2$O$_3$(0001) ($-0.8\sim4.0$ eV) originate from HgTe completely. As indicated by the orbital-resolved bands in Fig. \ref{sub-band}(c) and \ref{sub-band}(d), the Dirac-like bands that cross $E_F$ are contributed by the Hg$-s$ and Te$-p_{x,y}$ orbitals, while the parabolically dispersive band bellow $E_F$ is mainly from Te$-p_{x,y}$. These characters are similar to those in the freestanding honeycomb HgTe ML. In the absence of SOC, HgTe/Al$_2$O$_3$(0001) is a semimetal with a Dirac cone at $\vb\Gamma$ point. When the SOC effect is included, a large band gap of 227.2 meV opens at the Dirac point with$E_F$ locating in the gap. The SOC also mediates the interaction between the Hg$-s$ and Te$-p_{x,y}$ orbitals, leading to the band inversion near $E_F$ at $\vb\Gamma$ point, as shown in Fig. \ref{sub-band}(f) and \ref{sub-band}(g). Therefore, HgTe/Al$_2$O$_3$(0001) maintains the basic electronic features of the freestanding honeycomb HgTe ML, which implies the nontrivial band topology in the type-I HgTe/Al$_2$O$_3$(0001).

\begin{figure}
\centering
\includegraphics[width=8.5cm]{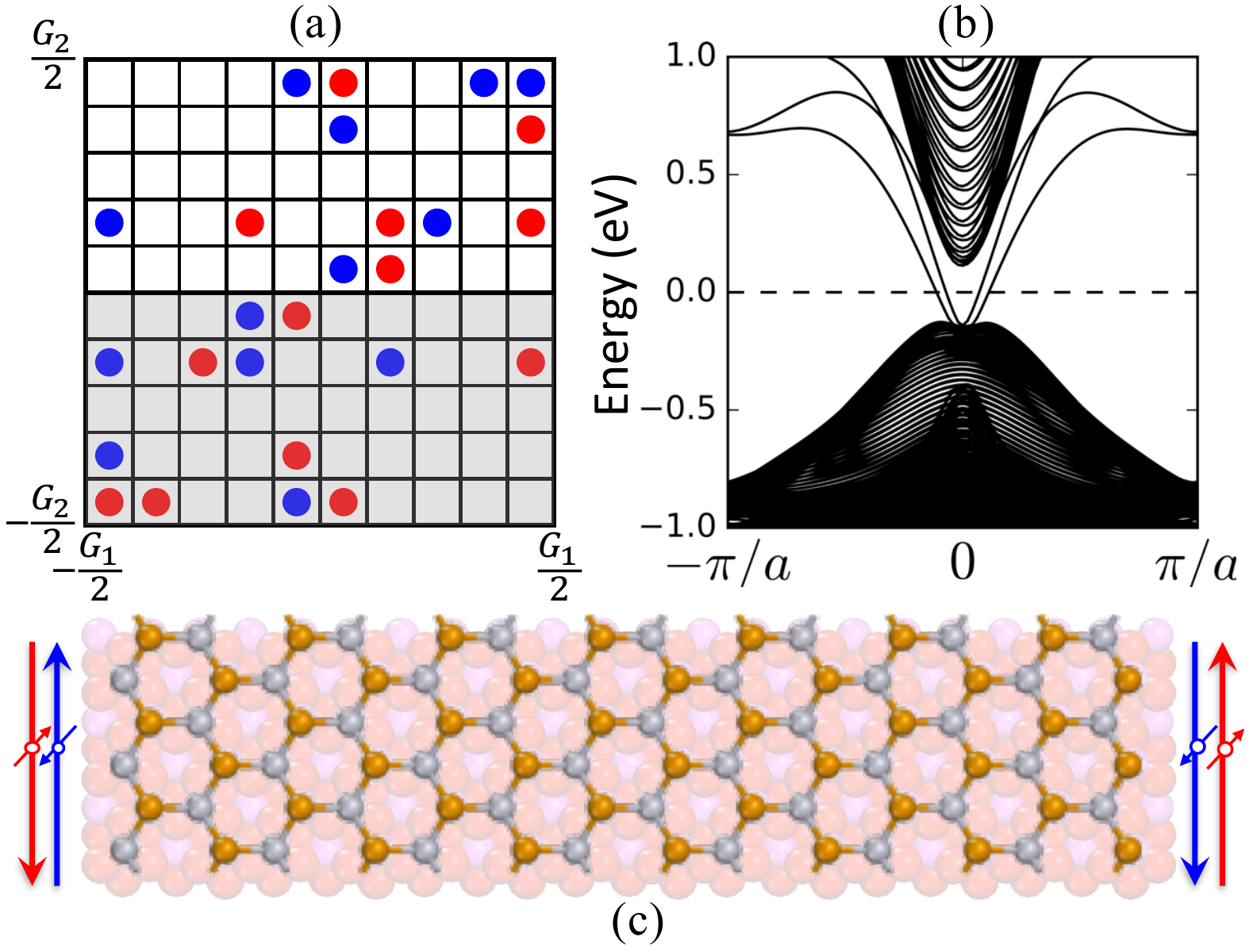}
\caption{(Color online) (a) The $n-$field configuration of the type-I HgTe/Al$_2$O$_3$(0001). The nonzero points are denoted by red ($n=1$) and blue ($n=-1$) dots, respectively. (b, c) Band structure and atomic structure of 1D type-I HgTe/Al$_2$O$_3$(0001) nanoribbon with zigzag edges. The actual width of the 1D nanoribbon in calculation is 50 $u.c.$. The horizontal dashed line represents the Fermi level. The long arrows in (c) indicate the directions of the edge currents. The short arrows denote the spin-up (red) and spin-down (blue) electrons, respectively.
}\label{sub-edge}
\end{figure}

To confirm the topological character of the type-I HgTe/Al$_2$O$_3$(0001), we calculated the $\mathbb{Z}_2$ invariant with the $n-$field method. From the $n-$field configure in Fig. \ref{sub-edge}(a), we obtained $\mathbb{Z}_2=1$, which is the clear evidence for nontrivial band topology. This can be further demonstrated by the band structure of the 1D nanoribbon of the type-I HgTe/Al$_2$O$_3$(0001) from tight-binding simulation as plotted in Fig. {sub-edge}(b). Clearly, helical edge states exist in the topological insulator gap, even through the dispersions of the edge bands are different from those of the freestanding HgTe nanoribbon, due to the change of the symmetry imposed by the substrate. Accordingly, the Al$_2$O$_3$(0001) substrate can not only support the honeycomb HgTe ML but also preserve the QSH states, which benefits the experimental investigations and practical applications. It is worth to note that the substrate of HgTe/Al$_2$O$_3$(0001) is a wide-band-gap insulator, so the edge currents propagate only along the edges of the honeycomb HgTe ML and can not leak into the substrate. Meanwhile, there are no free carriers in the substrate to disturb the edge currents. Therefore, the quantized edge currents are restricted within the HgTe ML and well protected against the electronic contamination from the substrate such as in the silicene/Ag(111) system. On the contrary, HgSe/Al$_2$O$_3$(0001) becomes a trivial insulator with band gap of 165 meV and $\mathbb{Z}_2=0$ (see Fig. S2). 

\begin{figure}
\centering
\includegraphics[width=8.5cm]{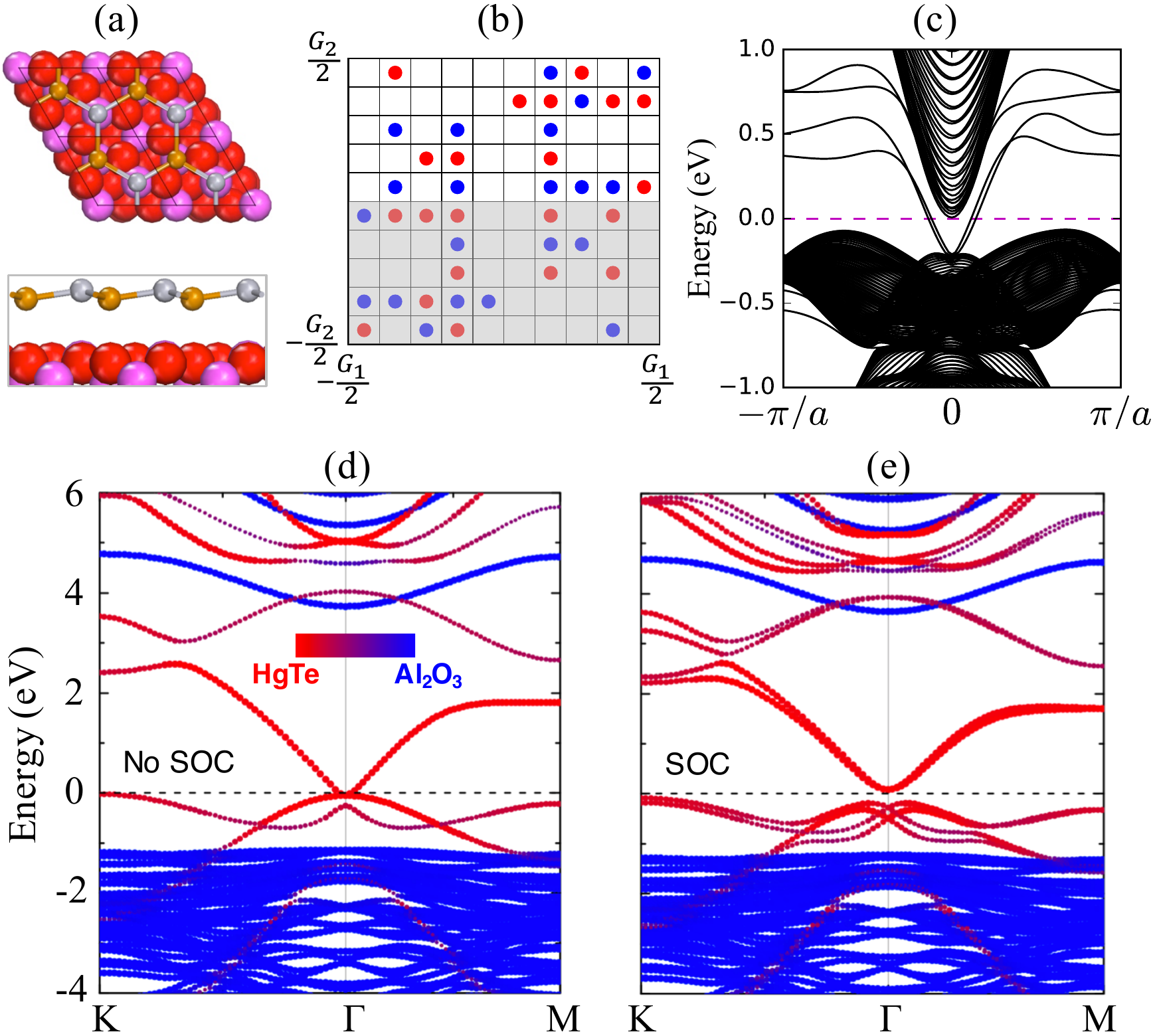}
\caption{(Color online) Properties of type-II HgTe/Al$_2$O$_3$(0001). (a) Top and side views of the atomic structure. (b) The $n-$field configuration in 2D Brillouin zone. The nonzero points are denoted by red ($n=1$) and blue ($n=-1$) dots, respectively. (c) Band structure of 1D nanoribbon with zigzag edges and width of 50 $u.c.$. (d, e) The band structures of the 2D periodic system without and with SOC. The horizontal dashed line represents the Fermi level.
}\label{type-II}
\end{figure}

%\section*{Discussion}
In the realistic growth environment, it is possible to get the type-II HgTe/Al$_2$O$_3$(0001) [Fig. \ref{type-II}(a)] in unequilibrium condition, although it has much smaller binding energy than that of the type-I configuration. Therefore, we calculated the corresponding band structures as displayed in Fig. \ref{type-II}(d) and \ref{type-II}(e). It is clear that the energy bands near $E_F$ are quite different from those of the freestanding honeycomb HgTe ML, because of the interaction between HgTe and Al$_2$O$_3$(0001), although the interaction is weak. Nevertheless, the band orders at the $\vb\Gamma$ point are the same as those of the type-I HgTe/Al$_2$O$_3$(0001) (see Fig. S3). Furthermore, the band structures clearly show that the SOC opens a band gap of 171 meV near $E_F$. Interestingly, this gap is also topologically nontrivial, which can be manifested by the nonzero $\mathbb{Z}_2$ invariant ($\mathbb{Z}_2=1$) in Fig. \ref{type-II}(b) and the quantized edge states in the bulk gap in Fig. \ref{type-II}(c). This is good for experimental realization and investigation of the QSH effect in HgTe/Al$_2$O$_3$(0001), because the exact configuration of the atomic structure is not a exclusive precondition to realize the nontrivial topological insulator states in HgTe/Al$_2$O$_3$(0001).

Another factor to hinder the realization of the QSH effect in HgTe/Al$_2$O$_3$(0001) is that the thickness of HgTe is more than one monolayer, which may ruin band topology. Therefore, we calculated the band structures of HgTe/Al$_2$O$_3$(0001) with two MLs of HgTe as shown in Fig. S4. We found that the type-I structure becomes metallic with two MLs of HgTe on Al$_2$O$_3$(0001) no matter the SOC effect is included or not. In contrast, the type-II structure maintains topologically nontrivial. However, the binding energy of the type-II structure is 0.78 eV larger than that of the type-I structure, so the possibility to get the type-I structure is much larger than the type-II structure. Therefore, the thickness of HgTe should be restricted to one monolayer to realize and investigate the QSH effect in HgTe/Al$_2$O$_3$(0001).

With the large topological insulator gap in the type-I HgTe/Al$_2$O$_3$(0001), it is expected to observe the QSH effect at high temperature. Therefore, it is important to investigate the thermal stability of the atomic structure of HgTe/Al$_2$O$_3$(0001). To this end, we carried out $ab~initio$ molecular-dynamics simulation at 300 K. We found that the binding between HgTe and Al$_2$O$_3$(0001) keeps strong, so that the HgTe ML does not detach or drift on Al$_2$O$_3$(0001). Furthermore, the HgTe monolayer maintains the honeycomb structure against the thermal vibration at 300 K (see Fig. S5). This suggests that HgTe/Al$_2$O$_3$(0001) can survive at room temperature and the QSH effect may be observed. Furthermore, the QSH state of HgTe/Al$_2$O$_3$(0001) may be robust under some other factors such as the moderate external electric field, because the insulating Al$_2$O$_3$(0001) substrate can effectively prevent further charge transfer between HgTe and Al$_2$O$_3$(0001). In contrast, the external electric field significantly affects the QSH state in CdTe/HgTe/CdTe quantum well. \cite{JAP-QW}

\section{Conclusion}

In summary, we proposed a robust QSH insulator which is composed of low buckled honeycomb HgTe monolayer and $\alpha-$Al$_2$O$_3$(0001), based on first-principles and tight-binding simulations. The analyses of the electronic structures reveal that HgTe/Al$_2$O$_3$(0001) is topologically nontrivial with nonzero topological invariant $\mathbb{Z}_2=1$. The topological insulator gap is as large as 227 meV. The band topology is caused by the band inversion between the energy levels of $\ket{s^-}$ and $\ket{p_{xy}^+}$ of the Hg and Te atoms. Interestingly, both the ground-state and metastable structures are topologically nontrivial and the atomic structures are stable at high temperature. Furthermore, the insulating substrate not only supports and stabilizes HgTe, but also protects the helical edge states against electronic contamination from substrate. These features are beneficial to experimental realization of the QSH effect at relatively high temperature. Our strategy paves the way for engineering realistic 2D TI which may be achieved in experiment and used for real applications at room temperature.

\addcontentsline{toc}{chapter}{Acknowledgment}
\section*{Acknowledgments}
This work is supported by the National Natural Science Foundation of China (11574223), the Natural Science Foundation of Jiangsu Province (BK20150303) and the Jiangsu Specially-Appointed Professor Program of Jiangsu Province.

%\section*{References}

\end{document}